\documentclass[aps,prl,twocolumn,showpacs,superscriptaddress,floatfix]{revtex4}
\usepackage{graphicx}
\usepackage{dcolumn}
\usepackage{bm}
\begin{document}
\title{Non Identical strange particle correlations\\
in Au-Au collisions at $\sqrt{s_{NN}} = 200$ GeV\\
from the STAR experiment\\}

\author{G. Renault}
 \email{gael.renault@subatech.in2p3.fr}
 \affiliation{SUBATECH,\\
Laboratoire de Physique Subatomique et des Technologies Associ\'ees\\ 
University of Nantes - IN2P3/CNRS - Ecole des Mines de Nantes \\
4 rue Alfred Kastler, F-44307 Nantes Cedex 03, France\\}
\author{The STAR Collaboration}
\noaffiliation{}
\pacs{25.75.Gz}

\begin{abstract}
{\bf Abstract} : Information about the space-time evolution of colliding nuclei
can be extracted correlating particles emitted from
nuclear collisions. The high density of particles produced in the STAR
experiment allows the measurement of non-identical strange
particle correlations.
Due to the absence of Coulomb interaction, $p-\Lambda$ and $\overline{p}-\Lambda$
systems are more
sensitive to the source size than $p-p$ pairs.
Strong interaction potential has been studied using
$p-\Lambda$, and for the first time, $\overline{p}-\Lambda$ pairs.
The experimental correlation functions have been described in
the frame of a model based on the $p-n$ interaction.
The first preliminary measurement of $\pi$ - $\Xi$
correlations has been performed, allowing to extract
information about the freeze-out time and the space-time asymmetries
in particle emission closely related to the transverse radial expansion
and decay of resonances.
\end{abstract}
\maketitle
\section{Introduction}\label{intro}

The Relativistic Heavy Ion Collider (RHIC) provides
the facility for colliding gold ions at 200 GeV 
per nucleon pair in the center of mass. 
The STAR detector (a Solenoid Tracker At RHIC), installed 
at RHIC collider, allows the reconstruction of the 
particles produced during the collisions. 
Non-identical particles
are correlated due to final state Coulomb and nuclear interactions.
So one can use 
the correlation technique to study the Final State
Interaction (FSI). 
In addition, non-identical particle correlations are sensitive to the space-time asymmetries
of the emission points of different particle species \cite{1}.

Contrary to $p-\Lambda$ \cite{2,5,6}, 
the nuclear FSI of $\overline{p}-\Lambda$
is still unknown.
In the following, data are shown and 
$p-\Lambda$ correlations are analysed
using the Lednick\'y \& Lyuboshitz model \cite{3}.

Preliminary results on $\pi \Xi$ correlations are
also shown and the dominated Coulomb FSI is observed.

\section{Experimental correlation functions}\label{xpcf}
Particles are measured in Au-Au collisions at 
 $\sqrt{s_{NN}} = 200$ GeV using the Time Projection Chamber
(TPC). Central events accounting for 10\% of the total 
cross section are selected.
Protons and anti-protons
are selected using their specific energy loss ($dE/dx$).
This selection limits the acceptance of particles to 
the transverse momentum range of 0.4-1.1 GeV/c in the 
rapidity interval $|Y|<0.5$.

The contamination and the feed-down have been studied
in order to estimate the purity ($\lambda$) of protons
as a function of the transverse momentum ($p_{t}$).
The purity is defined as the product of the probability of identification (Pid) 
times the fraction of primary protons (Fp).  
\begin{eqnarray}
\label{Purity}
\lambda (p_{t}) = \text{Pid} (p_{t}) * \text{Fp} (p_{t})
\end{eqnarray}
Values of purity indicated in the following correspond 
to the average transverse momentum of protons ($<p_{t}>=0.70$ GeV/c)
in the studied range.
The fraction of identified protons is estimated
to be 76.5\%.  
The feed-down study
leads to an estimated purity of 54\% for primary protons. 
Most of the secondary protons come
from lambda decay and represent 35\% of the protons
used to construct the correlation function. Other sources
of contamination of protons are provided by decay products 
of $\Sigma^{+}$ and 
pions interacting with matter, which represent respectively
10\% and 1\% of the sample.

The same study has been done for anti-protons ($<p_{t}>=0.73$ GeV/c),
times the fraction of identified anti-protons is estimated
to be 73\%.
The feed-down study
leads to an estimated purity of 56\% for primary anti-protons. 
Secondary protons are weak decay products or electromagnetic decay
products.
Most of the secondary anti-protons come
from anti-lambda decay and represent 32\% of the anti-protons
used to construct the correlation function.
For anti-protons, the additional source
of contamination, which is the decay products of $\overline{\Sigma^{+}}$,
represents 12\%.
Lambdas are reconstructed through the decay channel
$\Lambda \rightarrow \pi + p$, with a corresponding branching ratio
of 64\%.
Pions and protons are selected using their 
specific energy loss. In addition some geometrical cuts
are applied, giving a lambda purity sample of 80\%,
the remaining 20\% representing the combinatory background.
Only lambdas in the rapidity 
range $|Y|<1.5$ are selected.
Due to the acceptance of the detector the transverse momentum 
range is 0.3-2.0 GeV/c.
By reconstructing invariant mass of lambdas, 
we estimate misidentified lambdas as the combinatory background.
The sample of lambdas includes secondary particles such as decay products of
$\Xi, \Xi^{0}, \Sigma^{0}$.
Identified lambdas ($<p_{t}>=1.20$ GeV/c) represent 80\% of the particle sample used
to construct the correlation function.
Moreover an in-depth study of feed-down leads to 
the estimated lambda purity of 46\%.

Pair purity plays a crucial role in the correlation study.
The estimated value of the pair purity for $p-\Lambda$ and 
$\overline{p}-\Lambda$ systems is 15\%. 

$\Xi$ are reconstructed through the decay channels
$\Xi \rightarrow \Lambda + \pi$, with a corresponding 
branching ratio of 99.9\%.
$\Xi$ are not contaminated notably by any kind 
of particles. The $\Xi$ contamination due to $\Omega$
decay represents less than 3\%.
The study of particle and pair purity is under progress.
Pions are selected using their specific energy loss ($dE/dx$).

The contamination tends to reduce the correlation strength.
The non-correlated background is constructed by mixing events 
with primary vertex separated from each other by less than 10 cm.

The relevant variable is the momentum of one 
of the particle
in the pair rest frame called here 
$\overrightarrow{k}^*$.

\begin{figure}[htb!]
\center
\includegraphics[width=0.4\textwidth]{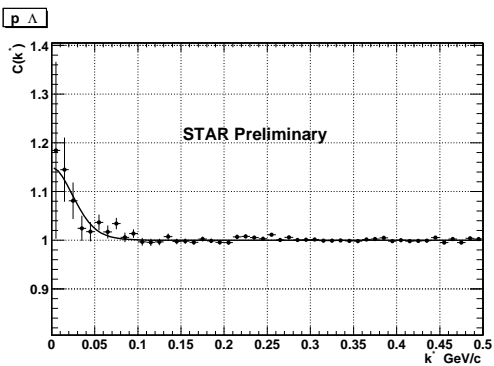}
\caption{The measured $p-\Lambda$ correlation function discribed by
Lednick\'y \& Lyuboshitz analytical model.} 
\label{pla}
\end{figure}

\begin{figure}[hbt!]
\center
\includegraphics[width=0.4\textwidth]{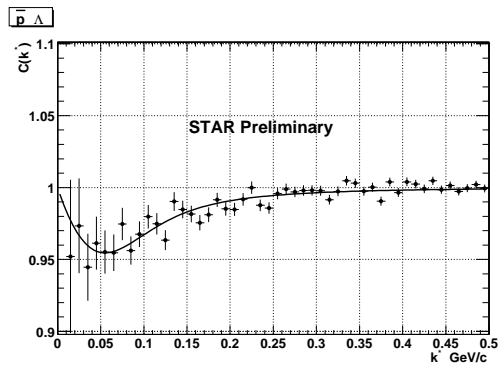}
\caption{The measured $\overline{p}-\Lambda$ correlation function by
Lednick\'y \& Lyuboshitz analytical model.} 
\label{pbarla}
\end{figure}

The correlation function has been extracted by 
constructing the ratio of two distributions.
The numerator is the modulus $\overrightarrow{k^{*}}$ distribution
of pairs from the same event. 
The denominator 
is the modulus $\overrightarrow{k^{*}}$ distribution of pairs composed 
of particles from different events. 
The $p-\Lambda$ correlation function is presented in 
Fig.\ref{pla}.
Fig. \ref{pbarla} represents the $\overline{p} \Lambda$
correlation function, which appears to be negative,
measured for the first time.

\section{ Lednick\'y \& Lyuboshitz analytical model}\label{model}

The analytical model used 
to study the correlation functions 
 is based on the $p-n$ 
final state interaction, described in the frame of the effective range approximation 
\cite{3,4} since only nuclear FSI is present in $p-\Lambda$ 
and $\overline{p}-\Lambda$ systems.
The correlation function ($C(k^{*})$)
is the average of the wave function ($\Psi^{s}$) 
over spin state (s) and over
the distribution of relative distance ($r^{*}$) of particles in the source:
\begin{eqnarray}
C(k^{*}) = \langle \left| \Psi^{s}_{- \vec{k^{*}}} 
(\vec{r^{*}})\right|^{ 2}  \rangle_{\vec{r^{*}},s}
\end{eqnarray}
The wave function is assumed to be equal to the leading term
s-wave, plus the scattering amplitude:
 
\begin{eqnarray}
\Psi_{- \vec{k^{*}}}^{s} (\vec{r^{*}}) = e^{-i\vec{k^{*}}.\vec{r^{*}}} 
+ \frac{f^{s}(k^{*})}{r^{*}} e^{-\vec{k^{*}}.\vec{r^{*}}}
\end{eqnarray}
with the scattering factor:
\begin{eqnarray}
f^{s}(k^{*}) = ( \frac{1}{f^{s}_{0}} 
+ \frac{1}{2} d^{s}_{0} k^{*2} - i k^{*} )^{-1}
\end{eqnarray}
where $f^{s}_{0}$ is the scattering length and
$d^{s}_{0}$ is the potential effective range, for the spin state (s), 
which is triplet (T) or singlet (S)
for the $p-\Lambda$ system.

The correlation function has the following expression:
\begin{eqnarray}
C(k^{*}) = \langle\left| e^{-\vec{k^{*}}.\vec{r^{*}}} 
+ \frac{f^{s}(k^{*})}{r^{*}} e^{-\vec{k^{*}}.\vec{r^{*}}} 
\right|^{ 2} \rangle_{\vec{r^{*}},s} 
+ O(\frac{1}{r_{0}^{3}})
\end{eqnarray}
The relative position distribution is assumed to be a gaussian distribution,
$r_{0}$ is the radius of the source.
\begin{eqnarray}
\vec{r^{*}} \sim e^{- \vec{ r^{*} }^{2} /4r_{0}^{2}}
\end{eqnarray}
We consider that particles are not polarized.
For $p-\Lambda$, the purity parameter is 
introduced as a suppression parameter 
to take into account the pair contamination.

For $\overline{p}-\Lambda$, the spin dependence is neglected
$f ^{S}  =  f^{T}  =  f$ and the effective range ($d_{0}$) is set to zero,
in order
to stronger constrain fitted parameters. Indeed, an extra parameter, 
$Im(f_{0})$ should be introduced 
to take into account 
the annihilation channel $B \overline{B}$.

The fit parameters from \cite{2} have been used 
for $p-\Lambda$ to extract values of the radius 
($r_{0}$) and the purity ($\lambda$)
(Table \ref{pLaParam}). Fits are stable 
with the values of parameters given in \cite{4}.
\begin{table}[hbt!]
\begin{tabular}{|c|c|c|c|c|}
 \hline
  Parameter & Value   \\ \hline
  $f^{S}_{0}$  (fm) & 2.88  \\ \hline 
  $d^{S}_{0}$  (fm) & 2.92  \\ \hline 
  $f^{T}_{0}$  (fm) & 1.66  \\ \hline 
  $d^{T}_{0}$  (fm) & 3.78  \\ \hline 
  $r_{0}$  (fm) & $2.71 \pm 0.33$  $\star$      \\ \hline
  $\lambda$ & $0.15$  \\ \hline
\end{tabular}
\caption{Parameters of $p-\Lambda$ 
interaction, $\star$  indicates
the fitted value.
Only statistical errors are indicated.
 $\lambda $ is the pair purity.}
\label{pLaParam}
\end{table}

The extracted source parameters are close to 
values obtained in measurements
performed by NA49 (CERN) collaboration in Pb+Pb collisions at 158 AGeV  \cite{5}
and by the E895 (AGS) experiment in Au+Au
collisions at $4$, $6$, and $8$ AGeV \cite{6}.

The radius extracted from $p-\Lambda$ correlation function,
is smaller than the one extracted from $\overline{p}-\Lambda$ correlation function. 

Using the correlation function allows
to estimate the final state interaction parameters for 
the $\overline{p}$ - $\Lambda$ system Table \ref{pbarLaParam}.
\begin{table}[hbt!]
\begin{tabular}{|c|c|c|c|c|}
 \hline
 Parameter & Value    \\ \hline
 Im($f_{0}$) (fm) & $1.21_\pm 0.94 $   $\star $      \\ \hline
 Re($f_{0}$) (fm) & $-2.65 \pm 1.26 $   $\star $      \\ \hline
 $d_{0}$ (fm) & 0.0	\\ \hline
 $r_{0}$ (fm) & $ 1.43 \pm 0.07$    $\star  $  \\ \hline
 $\lambda$ & 0.15 \\ \hline
\end{tabular}
\caption{Parameters of $\overline{p}-\Lambda$ 
correlation,
$\star$ indicates the fitted values, imaginary and real part of the scattering length and the radius. Only statistical errors
are shown. $\lambda $ is the pair purity.}
\label{pbarLaParam}
\end{table}
One can notice that the value of the imaginary part of the 
scattering length obtained for $\overline{p} \Lambda$ (Table \ref{pbarLaParam})
is in agreement with the scattering length ($0.8$ fm) 
for $\overline{p}-p$ spin averaged in \cite{7} .
\begin{figure}[hbt!]
\center
\includegraphics[width=0.4\textwidth]{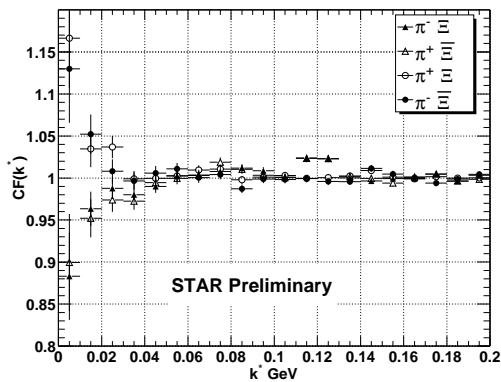}
\caption{$\pi-\Xi$ correlation functions.} 
\label{pixi}
\end{figure}

A preliminary study of $\pi-\Xi$ has been done (Fig. \ref{pixi}).
The dominant Coulomb FSI is visible for like-sign and unlike-sign
systems. The $\Xi^{*}(1530)$ resonance is observed at $k^{*} = 145 GeV/c$,
the shift from $152$ MeV/c \cite{8} to $145$ MeV/c is due to
a Lorentz boost to the pair rest frame. The peak for like sign particles, 
at $120$ MeV/c, is under investigation.\\
\\
\section{Conclusion}\label{conclusion}
Preliminary results on $p-\Lambda$, $\overline{p} \Lambda$ and $\pi-\Xi$ 
correlations have been shown. The analysis are still under progress.

It has been shown that by studying $p-\Lambda$ 
one can estimate the size of the
source of particles.

The surprising shape of the $\overline{p}-\Lambda$
correlation function
has been shown for the first time.

Final state interaction parameters, such
as the scattering length,
can be extracted from  $\overline{p}-\Lambda$.
The studied momentum resolution does not have any relevant effect.

The pair purity has a stronger effect on the correlation function
than momentum resolution
and is an important source of
systematic errors.
In addition the purity study is based on the assumption
that the contamination is
uncorrelated. Residual correlations may 
contribute to measured correlation and should be investigated.
The first $\pi-\Xi$ correlation functions show a dominant Coulomb FSI.

\end{document}